\documentclass[twocolumn,nofootinbib,amsmath,amssymb,a4paper]{revtex4}

\usepackage{graphicx}
\usepackage{dcolumn}
\usepackage{bm}

\begin{document}

\title{Measuring $CP$ violation in $B_{s}^{0}\rightarrow\phi\phi$ with LHCb}
\author{J. F. Libby 
on behalf of the LHCb Collaboration} \email{j.libby1@physics.ox.ac.uk}
\affiliation{%
    University of Oxford \\
}%

\begin{abstract}
Sensitivity studies to the $CP$-violating parameters of the decay $B_{s}^{0}\rightarrow\phi\phi$ with the LHCb 
experiment are presented. The decay proceeds via a $b\rightarrow ss\bar{s}$ gluonic-penguin quark 
transition, which is sensitive to contributions from beyond the Standard Model particles. A 
time-dependent angular analysis of simulated data leads to an expected statistical uncertainty of 
$6^{\circ}$ on any new physics induced $CP$-violating phase for a sample corresponding to 
2~$\mathrm{fb}^{-1}$ of integrated luminosity. The expected precision on $\sin 2\beta$ from the 
related decay  $B^{0}\rightarrow\phi K^{0}_S$ is also discussed.
\end{abstract}

\maketitle

\section{Introduction}
\begin{figure}
\includegraphics[height=0.4\textwidth,angle=270]{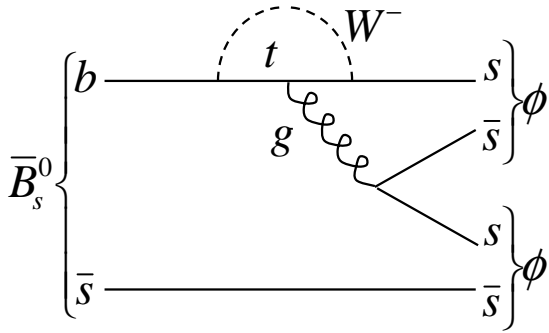}
\caption{\label{fig:feynbtophiphi} The main diagram contributing to the decay $B_{s}^{0}\rightarrow\phi\phi$.}
\end{figure}

The amplitude for the decay $B_{s}^{0}\rightarrow\phi\phi$ is dominated
by gluonic-penguin quark transitions $b\rightarrow ss\bar{s}$
(Fig.~\ref{fig:feynbtophiphi}). Gluonic-penguin processes are
sensitive to beyond the Standard Model particles that contribute within
the loop. The $e^{+}e^{-}$ $B$-factories have measured $\sin 2\beta$
in nine $B^{0}$ gluonic penguin decay modes, such as
$B^{0}\rightarrow\phi K^{0}_{S}$ and $B^{0}\rightarrow\eta^{\prime}
K^{0}_{S}$ \cite{bib:hfag}. All the measurements of $\sin 2\beta$
from these modes have values below that measured in $b\rightarrow
c\bar{c}s$ transitions, but no individual measurement shows a
significant deviation.

The decay $B_{s}^{0}\rightarrow\phi\phi$ is predicted to have a $CP$-violating 
phase less than $1^{\circ}$ within the
SM \cite{bib:raidal}. The dependence on $V_{ts}$ in both the mixing
and decay amplitudes leads to a cancellation of the
$B_{s}^{0}$-mixing phase. Therefore, if any significant $CP$-violation is measured in
$B_{s}^{0}\rightarrow\phi\phi$ decays it is an unambiguous signature of new
physics. The decay is of a pseudoscalar meson to two vector
mesons, which leads to the final state being a $CP$-even and $CP$-odd admixture.
Therefore, a time-dependent angular analysis is required to extract the $CP$-violating parameters 
of the decay.

The paper is organized as follows. Section~\ref{sec:lhcb}
contains a brief description of the LHCb experiment. The predicted
event yields and background estimations are described in
Section~\ref{sec:selection}. The $CP$ sensitivity study is presented
in Section~\ref{sec:cpstudy}. The LHCb prospects with the related
mode $B^{0}\rightarrow\phi K^{0}_{S}$ are discussed in
Section~\ref{sec:phikshort}. The conclusions are given in
Section~\ref{sec:conc}.

\section{The $\mathbf{LHCb}$ experiment}
\label{sec:lhcb} The Large Hadron Collider (LHC) collides protons at
a centre-of-mass energy of 14~TeV. The LHC produces $10^{12}$ $b\bar{b}$ quark pairs 
per nominal year of data-taking $(10^7~\mathrm{s})$ when operating at an 
instantaneous luminosity of $2\times 10^{32}~\mathrm{cm}^{-2}\mathrm{s}^{-1}$.\footnote{This luminosity optimises the number of single interactions 
per bunch crossing.} 
The LHCb spectrometer \cite{bib:tp,bib:reopttdr} instruments 
one forward region about the $pp$ collision point. The forward geometry captures
approximately one-third of all $B$ hadrons produced and increases the
probability of both $B$ hadrons from the $b\bar{b}$ pairs
being within the acceptance, which improves the efficiency of
flavour tagging.

A silicon vertex detector, with sensors perpendicular to the beam
axis, is situated close to the interaction region in a secondary
vacuum. The detector provides accurate determination of primary and
secondary vertices leading to a proper-time resolutions of
approximately 40~fs in hadronic $B$-decays such as $B_{s}^{0}\rightarrow\phi\phi$. Tracking stations either side of a 1.2~T dipole
magnet produce momentum measurements with an accuracy of a few parts
per mille. There are two Ring-Imaging \v{C}erenkov detectors, with 3
different radiators, that allow identification of $K^{\pm}$ from
$\pi^{\pm}$ over the momentum range 1 to $100~\mathrm{GeV}/c$.

 In addition, the detector includes an electromagnetic calorimeter, a hadron
calorimeter and a muon detector. These components are critical for identifying
large transverse momentum, $p_{T}$, electrons, photons, hadrons and muons from
$B$-hadron decay in the initial hardware stage (Level-0) of the LHCb
trigger. The Level-0 trigger reduces the 40~MHz collision rate to
1~MHz. All data is then transferred from the detector to a dedicated
CPU farm where the Higher Level Trigger (HLT) algorithms are
performed. Initially an association between the high $p_{T}$ objects
that satisfy the Level-0 trigger and tracks with large impact
parameter to the primary vertex is sought. If this is successful,
exclusive and inclusive HLT algorithms are executed resulting in a
2~kHz output rate to disk.

\section{Event selection and background estimation}
\label{sec:selection} The estimation of $B_{s}^{0}\rightarrow\phi\phi$ yields and background has been 
performed on simulated data. The simulation of the $pp$ collisions and subsequent hadronisation is 
performed by the {\tt PYTHIA} generator \cite{bib:pythia}. The decay of any $B$ hadrons produced is
simulated by {\tt EVTGEN} \cite{bib:evtgen} and the detector response is performed by
{\tt GEANT4} \cite{bib:geant4}. The resulting data are processed by the 
complete LHCb reconstruction software. A dedicated sample containing events with a
$B_{s}^{0}\rightarrow\phi\phi$ decay is used to evaluate the selection efficiency.
Minimum bias events are selected rarely by the Level-0 trigger and the HLT.
Therefore, an inclusive sample of 34 million events containing $b\bar{b}$ quark
pairs is used to estimate the background. Due to the very large
number of $b\bar{b}$ pairs that will be produced at LHCb, the
inclusive $b\bar{b}$ sample only corresponds to approximately
15~minutes of data taking at the nominal instantaneous luminosity;
this leads to large uncertainties on the background estimates.

The selection process begins with the identification of $\phi$
candidates reconstructed from two oppositely charged kaons. Particle
identification and $p_T$ requirements are placed on the $K^{\pm}$
and they must be consistent with production at a common
vertex. The mass of the $K^{+}K^{-}$ must be within 20~$\mathrm{MeV/c^{2}}$ 
of the $\phi$ mass; the mass interval
corresponds to approximately $\pm 3\sigma_{m_{\phi}}$, where
$\sigma_{m_{\phi}}$ is the mass resolution.

\begin{figure}
\includegraphics[width=0.45\textwidth]{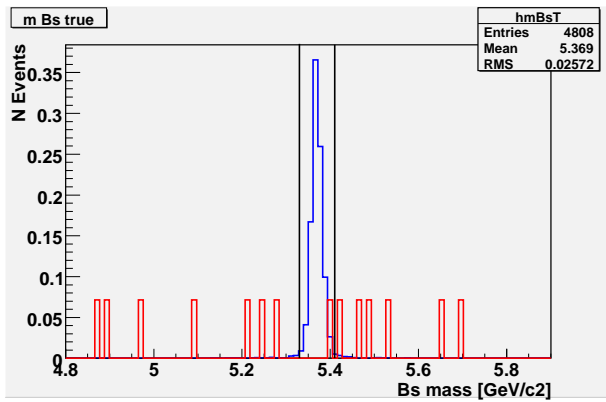}
\caption{\label{fig:Bmass} The $B_{s}^{0}$ mass of candidates reconstructed in 
the signal (blue) and inclusive $b\bar{b}$ simulation samples (red) before trigger selections. The normalisations are arbitrary.}
\end{figure}

$B$-meson candidates are reconstructed in events with two or more
$\phi$ candidates with a $p_T>1.2~\mathrm{GeV/c^{2}}$. The $B$-meson
candidates must be consistent with production at a common vertex
and this vertex must be well separated from the primary vertex.  The $B_{s}^{0}$
candidate mass distribution is shown in Fig.~\ref{fig:Bmass}
for signal and background samples. Events within 40
$\mathrm{MeV/c^{2}}$ of the $B_{s}^{0}$ mass are considered as
signal; the mass interval corresponds to approximately $\pm
3\sigma_{m_{B}}$, where $\sigma_{m_{B}}$ is the mass
resolution. Once Level-0 and HLT trigger selections have been applied there are 4000
signal events expected in every $2~\mathrm{fb}^{-1}$ of integrated
luminosity, which corresponds to one nominal year of LHC operation. The
event yield is calculated assuming the measured branching fraction 
$\mathcal{B}(B_{s}^{0}\rightarrow\phi\phi)=(14^{+6}_{-5}(stat.)\pm 6(syst.))\times 10^{-6}$ 
\cite{bib:cdfphiphi}; the measured value lies within
theoretically predicted range \cite{bib:phiphitheory1,bib:phiphitheory2}. 
The background remaining in the $b\bar{b}$ inclusive simulation sample is found to 
consist of combinatoric $B_{s}^{0}$ candidates. The background-to-signal 
ratio is bounded to lie between 0.4 to 2.1 at the 90\% confidence level, with 
a central value of 0.9. 

\section{$B_{s}^{0}\rightarrow\phi\phi$ $CP$ sensitivity}
\label{sec:cpstudy}

\begin{figure}[t]
\includegraphics[height=0.45\textwidth,angle=270]{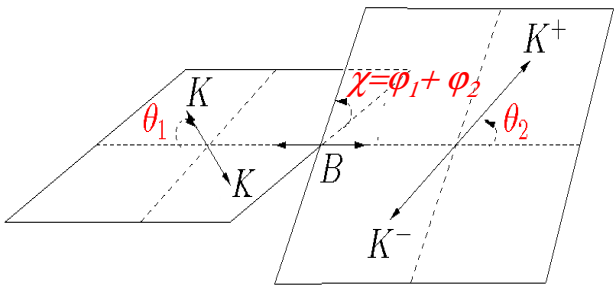}
\caption{\label{fig:tranang} A schematic of the definition of the 
transversity angles $\theta_1$, $\theta_2$ and $\chi$.}
\end{figure}

The magnitude of any new physics induced $CP$-phase, $\phi_{NP}$, is extracted 
from a time-dependent analysis of the differential cross section with respect to 
the three transversity angles defined in Fig~\ref{fig:tranang}.
The amplitude for the decay can be written in terms of three helicity amplitudes 
$H_{\lambda}(t)$ where $\lambda = 0,\pm 1$. The helicity amplitudes are related to the 
transversity basis by $A_{0}(t)=H_{0}(t)$, $A_{||}(t)=\frac{1}{\sqrt{2}}(H_{+1}+H_{-1})$ and 
$A_{\perp}(t)=\frac{1}{\sqrt{2}}(H_{+1}-H_{-1})$. The amplitudes $A_{0}$ and $A_{||}$ are $CP$ even and 
$A_{\perp}$ is $CP$ odd. The differential cross section is then given by:
\begin{widetext}
\begin{eqnarray*}
\label{eqn:diffxsec}
\frac{d\Gamma(t)}{d\chi d\cos{\theta_{1}}d\cos{\theta_{2}}} &=&
|A_{0}(t)|^{2}f_{1}(\chi,\theta_{1},\theta_{2})
|A_{||}(t)|^{2}f_{2}(\chi,\theta_{1},\theta_{2})+
|A_{\perp}(t)|^{2}f_{3}(\chi,\theta_{1},\theta_{2}) \\
&+& 
\Im(A_{0}^{*}(t)A_{\perp}(t))f_{4}(\chi,\theta_{1},\theta_{2})+
\Re(A_{0}^{*}(t)A_{||}(t))f_{5}(\chi,\theta_{1},\theta_{2})+
\Im(A_{||}^{*}(t)A_{\perp}(t))f_{6}(\chi,\theta_{1},\theta_{2}) \; (1),
\end{eqnarray*}
\end{widetext}
where $f_{i}\;(i=1-6)$ are even angular functions as required by Bose symmetry \cite{bib:phiphilhcb}. 
The time-dependent factors of these angular functions are sensitive to any 
$CP$-violating phase. In principle the $CP$-violating phase can be different between the three transversity amplitudes. However, to simplify the analysis it has been assumed to be equal. Significant fine-tuning of the phases would be required for no effects to be observed if new physics induced phases are present. The time-dependent terms also depend on the strong phase differences $\delta_{||,0}$ 
between $A(t)_{\perp}$ and $A(t)_{||,0}$, the relative magnitudes of the three 
amplitudes and the mass (lifetime) differences between the $B_s^{0}$ mass eigenstates, $\Delta m_s~(\Delta\Gamma_s)$.   

Simulated data are generated to follow the differential distribution given in Eqn.~\ref{eqn:diffxsec} with the value of $\phi_{NP}$ chosen to be 0.2~rad. 
The strong phases are assumed to be $\delta_{||}=0$ and $\delta_{0}=\pi$; these values are motivated by na\"{i}ve factorization \cite{bib:factorization}. 
The magnitudes of the transversity amplitudes are set to the values measured in the analogous  channel  for $B^{0}$ decays $B^{0}\rightarrow K^{*}\phi$ 
\cite{bib:babarkstphi,bib:bellekstphi}.
The value of $\Delta m_{s}$ is taken to be $17~\mathrm{ps}^{-1}$ \cite{bib:deltamscdf} and $\Delta\Gamma_s/\Gamma$ is set to be 0.15, compatible with current experimental constraints \cite{bib:bslifediffexp} and 
theoretical expectations \cite{bib:bslifedifftheory}. The signal sample size has a mean of 4000 events corresponding to an integrated luminosity of $2~\mathrm{fb}^{-1}$. 

\begin{figure}[t]
\includegraphics[height=0.45\textwidth,angle=270]{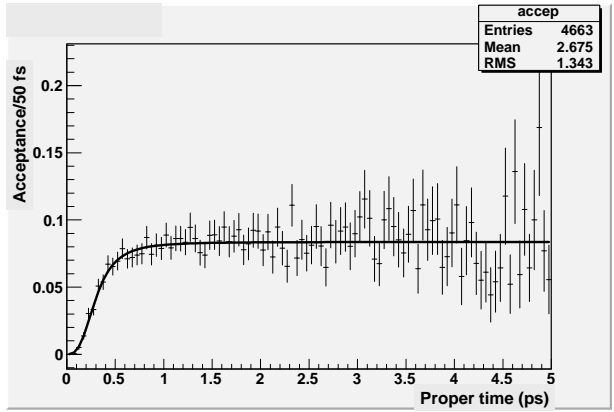}
\caption{\label{fig:lifetimeacc} The variation of the acceptance as a function of the proper-time, $\tau$. The acceptance function fit to the simulated 
data is $\epsilon(\tau)=\frac{0.084\tau^{3}}{0.027+\tau^{3}}$.}
\end{figure}

The following experimental effects are also simulated.
\begin{itemize}
\item{The {\bf background} is assumed to be at level 90\% of the signal with flat transversity angle and mass distributions, and an exponential 
lifetime distributions.} 
\item{The {\bf tagging power}, $\epsilon(1-2\omega)$, where $\epsilon$ is the tagging efficiency and $\omega$ is the mistag rate, is assumed to
be 9\% which has been found in simulation studies of other $B_{s}^{0}$ hadronic decays \cite{bib:lhcbtagging}. This is significantly better than that for $B^{0}$ modes because the kaon associated with 
the $B_{s}^{0}$ hadronisation is also used.}  
\item{The {\bf proper-time acceptance} measured from the signal simulation sample is shown in Fig. \ref{fig:lifetimeacc}. The reduced acceptance for
short lifetimes is the result of trigger and selection requirements on the impact parameters of the $B$ daughters.} 
\item{{The \bf proper time and $\mathbf{B_s^{0}}$ mass resolutions} are estimated to be 40~fs and $12~\mathrm{MeV}/c^{2}$, respectively. These resolutions are estimated from the signal simulation sample used for the selection studies.} 
\item{The {\bf angular acceptance} and {\bf resolution} are assumed to be flat and to have negligible effect, respectively; 
this assumption is motivated by the studies of related channel $B_{s}^{0}\rightarrow J\psi\phi$ \cite{bib:lhcbjpsiphi}.}
\end{itemize}

\begin{figure}[pth]
\includegraphics[height=0.45\textwidth,angle=270]{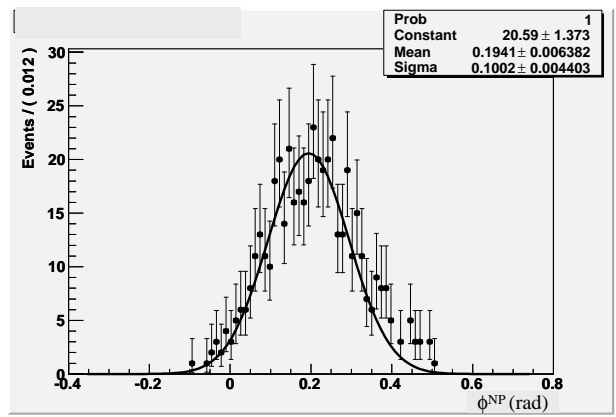}
\caption{\label{fig:fitresphiphi} The distribution of 
fitted value of $\phi_{NP}$ for 500 simulated $B_s^{0}\rightarrow\phi\phi$ 
experiments.}
\end{figure}

Five-hundred samples of signal and background events are generated and 
a maximum likelihood fit is performed on each one. The parameters $\phi_{NP}$, $\delta_{||}$, $\delta_{0}$ and the magnitude of the transversity amplitudes 
are extracted from the fit; all other parameters are fixed. 
The distribution of the fitted value of $\phi_{NP}$ for these 500 experiments 
is shown in Fig \ref{fig:fitresphiphi}. 
The average error on $\phi_{NP}$ is 0.1~rad ($5.7^{\circ}$). The pull distribution of the true value subtracted from the fitted value divided by the 
uncertainty is normal. 

Sets of 500 experiments are produced varying the input parameters assumed.
The variation of the uncertainty on $\phi_{NP}$ as a function of the $\mathcal{B}(B_s^{0}\rightarrow\phi\phi)$, signal-to-background ratio and $\Delta\Gamma_{s}/\Gamma_s$ is given in Table~\ref{tab:robust}. The variation of the assumed $\mathcal{B}$ leads to the expected statistical scaling of the uncertainty. 
The uncertainty on $\phi_{NP}$ is not degraded significantly until the background-to-signal ratio is greater than three. 
Increasing the value of $\Delta\Gamma_{s}/\Gamma$ leads to a 
reduction in the uncertainty because enhanced 
interference from the lifetime difference 
among the amplitudes increases the sensitivity to $\phi_{NP}$. The values of $\phi_{NP}$, the proper-time resolution and the relative magnitudes of the transversity amplitudes are also varied; all these had negligible effect on the sensitivity to $\phi_{NP}$.

\begin{table}[t]
\begin{center}
\caption{\label{tab:robust} 
The variation of the uncertainty on $\phi_{NP}$ as a function 
of the branching fraction, background-to-signal ratio (B/S) 
and $\Delta\Gamma_{s}/\Gamma_{s}$.}
\begin{tabular}{cc}\hline\hline
$\mathcal{B}~(\times 10^{-5})$ & $\sigma(\phi_{NP})$ \\ \hline
0.35 & $13^{\circ}$ \\
0.7  & $8.1^{\circ}$ \\
1.4  & $5.7^{\circ}$ \\ 
2.1  & $4.6^{\circ}$ \\ \hline\hline
B/S  & $\sigma(\phi_{NP})$ \\ \hline
0    & $5.5^{\circ}$ \\
0.9  & $5.7^{\circ}$ \\ 
2    & $6.1^{\circ}$ \\ 
5    & $7.2^{\circ}$ \\ \hline\hline
$\Delta\Gamma_s/\Gamma_s$ & $\sigma(\phi_{NP})$  \\ \hline
0.05 & $7.2^{\circ}$ \\
0.15 & $5.7^{\circ}$ \\
0.05 & $4.9^{\circ}$ \\ \hline 
\end{tabular} 
\end{center}
\end{table}

\section{$B^{0}\rightarrow\phi K^{0}_{S}$ CP sensitivity}
\label{sec:phikshort}
Simulation studies of the decay $B^{0}\rightarrow\phi K^{0}_{S}$ have also been
performed. The expected yield per $2~\mathrm{fb}^{-1}$ of integrated luminosity is 800 events.
These yields do not include $K^{0}_S$ daughters without measurements in 
the silicon vertex detector, approximately two-thirds of $K^{0}_{S}$ from these decays, which are not reconstructed in the current HLT algorithms. Algorithms to perform this reconstruction are currently being developed. The background-to-signal ratio is estimated to be 2.4 from the $b\bar{b}$ inclusive simulation. 

A time-dependent analysis of the $B^{0}\rightarrow\phi K^{0}_{S}$ is required to 
extract the sensitivity to $\sin{2\beta}$. 
A toy simulation study is performed to extract the sensitivity to $\sin{2\beta}$. The tagging power is assumed to be $5\%$ and the proper-time resolution is 
taken to be 60~fs from the simulated signal sample. A 10\% $K^{+}K^{-}$ $S$-wave contribution is also included in the fit. The uncertainty on $\sin{2\beta}$ is expected to be $0.32$ for a data sample corresponding to 
$2~\mathrm{fb^{-1}}$ of integrated luminosity.      

\section{Conclusions}
\label{sec:conc}
Sensitivity studies to $CP$-violation in the decay $B_{s}^{0}\rightarrow\phi\phi$ have been presented. A sample of data corresponding to an integrated luminosity of $2~\mathrm{fb}^{-1}$ gives an uncertainty of $6^{\circ}$ on any new physics induced $CP$ phase. Varying the assumptions used within reasonable ranges changes the predicted statistical uncertainty between $4^{\circ}$ and $13^{\circ}$. The largest statistical uncertainty results from decreasing the branching fraction by a factor of four. The measurement of $\sin{2\beta}$ from the $B^{0}\rightarrow\phi K^{0}_{S}$ has also been investigated. The sensitivity is expected to be 0.32 with a data set corresponding to $2~\mathrm{fb}^{-1}$ of integrated luminosity; this is of the same order as the current sensitivity of the $e^{+}e^{-}$ $B$-factories \cite{bib:bellephiks, bib:babarphiks}. Therefore, in conclusion, $B_s^0\rightarrow\phi\phi$ is the most sensitive mode with which to study gluonic-penguin $B$ decays with LHCb.


\begin{thebibliography}{99}
\bibitem{bib:hfag}
E. Barberio {\it et al.} (Heavy Flavour Averaging Group), {\tt 
http://www.slac.stanford.edu/xorg/hfag} \;. 
\bibitem{bib:raidal}
M. Raidal, Phys. Rev. Lett {\bf 89}, 231803 (2002).
\bibitem{bib:tp}
LHCb Technical Proposal, LHCb Collaboration, CERN/LHCC 98-4 (1998).
\bibitem{bib:reopttdr}
LHCb Reoptimized Detector Design and Performance, LHCb Collaboration, CERN/LHCC 2003-040 (2003).
\bibitem{bib:pythia}
T. Sj\"{o}strand, L. L\"{o}nnblad and S. Mrenna, hep-ph/0108264.
\bibitem{bib:evtgen}
D. Lange, Nucl. Instr. and Meth. A462 (2001) 152.
\bibitem{bib:geant4}
S. Agostinelli {\it et al.} (GEANT4 Collaboration), Nucl. Instr. and Meth. A506 (2003) 250.
\bibitem{bib:cdfphiphi}
D. Acosta {\it et al.} (CDF Collaboration), Phys. Rev. Lett. {\bf 95}, 031801 (2005).
\bibitem{bib:phiphitheory1} 
Y.-H.~Chen {\it et al.}, Phys. Rev. D {\bf 59}, 074003 (1999).
\bibitem{bib:phiphitheory2}
X. Q. Li, G. R. Lu and Y. D. Yang, Phys. Rev. D {\bf 68}, 114015 (2005) 
[Erratum-ibid D {\bf 71}, 019902 (2005)]. 
\bibitem{bib:phiphilhcb}
B. de Paula {\it et al.}, LHCb-2007-047.
\bibitem{bib:factorization}
M.~Bauer, B.~Stech and M.~Wirbel, Z. Phys. C {\bf 34}, 103 (1987).
\bibitem{bib:babarkstphi}
B.~Aubert {\it et al.} (BABAR Collaboration), Phys. Rev. Lett. {\bf 93} 231804 (2004).
\bibitem{bib:bellekstphi}
K.-F.~Chen {\it et al.} (Belle Collaboration), Phys. Rev. Lett. {\bf 94} 221804 (2005).
\bibitem{bib:deltamscdf} 
A.~Abulencia {\it et al.} (CDF Collaboration), Phys. Rev. Lett. {\bf 97} 242003 (2006).
\bibitem{bib:bslifediffexp} G.~Weber, these proceedings.
\bibitem{bib:bslifedifftheory} C.~Tarantino, hep-ph/0702235. 
\bibitem{bib:lhcbtagging} M.~Calvi, O.~Leroy and M.~Musy, LHCb-2007-058.
\bibitem{bib:lhcbjpsiphi} L.~Fernandez, CERN-THESIS-2006-042. 
\bibitem{bib:bellephiks} K.-F.~Chen {\it et al.} (Belle Collaboration), hep-ex/0608039.
\bibitem{bib:babarphiks} B.~Aubert {\it et al.} (BABAR Collaboration), hep-ex/0607112.  




\end{thebibliography}
\end{document}